\documentclass[12pt]{article}
\usepackage{amsfonts}
\usepackage{amssymb}

\begin{document}
\hspace{12.5cm}{\textbf{ETH-TH/97-29}}

\hspace{12.5cm}{September 1997}

\vspace*{1cm}
\centerline{\large{\bf{On the Construction of Zero Energy States}}}
\centerline{\large{\bf{in Supersymmetric Matrix Models}}}
\vspace{1cm}
\centerline{Jens Hoppe}
\centerline{Theoretische Physik, ETH H\"onggerberg}
\centerline{8093 Z\"urich, Switzerland}

\vspace{1cm}
\centerline{\large Abstract}
\vspace{0.5cm}
For the SU(N) invariant supersymmetric matrix model related to membranes in 4
space-time dimensions, the general solution to the equation(s)
$Q^{\dagger}\Psi=0$ $(Q\chi =0)$ is determined for N odd. For any such
(bosonic) solution of $Q^{\dagger}\Psi=0$, a 
(fermionic) $\Phi$ is found that (formally) satisfies
$Q^{\dagger}\Phi=\Psi$. 

For the analogous model in 11 dimensions the solution of $Q^{\dagger}\Psi=0
(Q\Psi=0)$ is outlined.

\vfill\eject  
Previous methods to study the existence of zero
energy bound states in SU(N)-invariant supersymmetric matrix models do not
lead to actual solutions of $(Q^{\dagger}Q+QQ^{\dagger})\Psi=0$. It therefore
seems valuable to propose a different route. For the model corresponding to 4
space time dimensions, I solve the equation(s) $Q^{\dagger}\Psi=0$ $(Q\chi
=0)$ and for any such $\Psi \ \epsilon \ {\cal H}_{+}$ (= the space of
bosonic, SU(N) invariant wavefunctions) determine a $\Phi$ satisfying
$Q^{\dagger}\Phi=\Psi$. By proving that $\Phi$ is normalizable, provided
$\Psi$ is, one could (as observed by J.~Fr\"ohlich) prove rigorously the
absence of a (bosonic) groundstate 
in this model, for arbitrary (odd) N. In cases when a ground state does exist
(as is expected, e.g., for the 11-dimensional model; see [1-9] for some literature), comparing the general solution of
$Q^{\dagger}\Psi=0$ with the 
general solution of $Q\chi =0$
(or proving $\Phi$ to be non-normalizable for some normalizable $\Psi=Q^{\dagger}\Phi$) will yield the explicit construction of the ground
state wave-function. For the supersymmetric matrix model corresponding to
membranes in 11 space-time dimensions, the calculation is set up in a form
where the determination of the expected ground state seems feasible.

Let
\begin{eqnarray}
Q   & = & 2\partial_a\frac{\partial}{\partial\lambda_a} + iq_a\lambda_a
\nonumber\\ 
Q^{\dagger} & = & -2\overline{\partial}_a\lambda_a - i q_a
\frac{\partial}{\partial\lambda_a}
\end{eqnarray}
where $\partial_a=\frac{\partial}{\partial z_a}, z_a \ \epsilon \ {\mathbb C},
q_a=\frac{i}{2} f_{abc}z_b\overline{z}_c$ ($f_{abc}$ being totally
antisymmetric, real, structure constants of SU(N)) and $\lambda_a$
$(\frac{\partial}{\partial\lambda_a})$ being fermionic creation (annihilation)
operators satisfying $\{\lambda_a, \
\frac{\partial}{\partial\lambda_b}\}=\delta_{ab} \ , \ \{\lambda_a ,
\lambda_b\}=0=\{\frac{\partial}{\partial\lambda_a} \ , \
\frac{\partial}{\partial\lambda_b}\}$. The hamiltonian of the model is
\begin{equation}
QQ^{\dagger}+Q^{\dagger}Q
=-4 \partial_a\overline{\partial}_a \ + \
q^2+f_{abc}z_c\lambda_a\lambda_b+f_{abc} \overline{z}_c\frac{\partial}{\partial\lambda_b}
\frac{\partial}{\partial\lambda_a}.
\end{equation}
$Q$ and $Q^{\dagger}$ commute with the operators of SU(N),
\begin{equation}
J_a \ := \
-if_{abc}(z_b\partial_c+\overline{z}_b\overline{\partial}_c+
\lambda_b\partial_{
  \lambda c}) 
\end{equation}
and on the Hilberspace ${\cal H}$ of gauge-invariant states,
$Q^2=i\overline{z}_aJ_a=0$.
Let
\begin{equation}
\Psi=\sum\limits^{N^2-1}_{j=0}
\frac{1}{j!} \ \psi_{a{_1}}\cdots_{a{_{2j}}}\lambda_{a{_1}}\cdots
\lambda_{a{_{2j}}} 
\end{equation}
(an analogous discussion could be applied to states in ${\cal H}_-$,
i.e. states containing only odd numbers of $\lambda$'s).
The equations $Q\chi=0, Q^{\dagger}\Psi=0$ then read 
\begin{equation}
i(2k-1) \ q[_{a{_{1}}}\chi_{a{_{2}}}\cdots_{a{_{2k-1}}}] \ = \ 2\partial{_{a}}
\chi_{a{_{1}}}\cdots_{a_{2k-1}a}
\end{equation}

\begin{equation}
(2k-1)2\overline{\partial}[_{a_{1}}\psi_{a{_{2}}}\cdots_{a{_{2k-1}}}] \ = \
iq_a\psi_{a{_1}}\cdots_{a{_{2k-1}a}} 
\end{equation}
where $k=1, \cdots , K:=\frac{N^2-1}{2}$.
Think of (5) as equations for
$\chi^{(2k-2)}=\{\chi_{a{_1}}\cdots_{a_{2k-2}}\}$, provided
$\chi^{(2k)}=\{\chi_{a{_2}}\cdots_{a_{2k}}\}$ is known; it is not 
difficult to verify that 
\begin{equation}
\chi^{[in]}_{a{_1}}\cdots_{a{_{2k-2}}}:= \frac{-2i}{q^2}
q_{a{_{2k-1}}}\partial_{a{_{2k}}} \chi_{a{_1}}\cdots_{a{_{2k}}}
\end{equation}
solves (5). At each stage of the interaction (eventually leaving only one
single free function,
$\chi_{a{_1}}\cdots_{a{_{2k}}}=\epsilon_{a{_1}}\cdots_{a{_{2k}}}\tilde{\chi}$)
a solution of the homogeneous equation, i.e.
\begin{equation}
q[_{a}\chi^{[h]}_{a{_{1}}}\cdots_{a{_{2k-2}}}] \ \equiv \ 0
\end{equation}
may be added. Similarly, one may verify that 
\begin{equation}
\psi^{(in)}_{a{_1}}\cdots_{a{_{2k}}}:=\frac{(2k)(2k-1)}{q^2} \ 2i \
q[_{a{_1}}\overline{\partial}_{a{_2}}\psi_{a{_3}}\cdots_{a{_{2k}}}]
\end{equation}
(determining $\psi^{(2k)}$ in terms of $\psi^{(2k-2)}$) solves (6), so that
the \textit{general} solution of (6) is given by
\begin{equation}
\Psi \ = \ \Psi^{(in)} \ \oplus \ \Psi^{(h)}
,\end{equation}
with
\begin{equation}
q_{a{_{2k}}} \ \psi^{(h)}_{a{_1}}\cdots_{a{_{2k}}} \ \equiv \ 0
.\end{equation}
With the direct sum property indicated in (10)
$(\int\psi^{(h)\star}_{a{_1}}\cdots_{a{_{2k}}} \
\psi^{(in)}_{a{_1}}\cdots_{a{_{2k}}} \ = \ 0 \ = \ \int
\psi^{[in]\star}_{a{_1}}\cdots_{a{_{2h}}}\psi^{[h]}_{a{_1}}\cdots_{a{_{2k}}})$
the choice of the particular solution(s) of the inhomogenous equation(s) is
canonical. 
Note that
\begin{equation}
(q\partial)\psi_{a{_1}}\cdots_{a{_n}} \ + \
n\psi_{a[a{_2}}\cdots_{a{_n}}\partial_{a{_1}]}q_{a}=0
\end{equation}
when $J_a\Psi=0$.
Now define
\begin{eqnarray}
\Phi=\sum\limits^K_{k=1}\frac{1}{(2k-1)!} \ \phi_{a{_1}}\cdots_{a{_{2k-1}}},
\lambda_{a{_1}}\cdots\lambda_{a{_{2k-1}}} \nonumber
\end{eqnarray}
by
\begin{equation}
\phi_{a{_1}}\cdots_{a{_{2k-1}}} \ := \ \frac{-i}{q^2} \ (2k-1) q \
[_{a{_1}}(\psi_{a{_2}}\cdots_{a{_{2k-1}}}]-2 \ (2k-2) \
\overline{\partial}_{a{_2}}\phi_{a{_3}}\cdots_{a{_{2k-1}}}]) 
\end{equation}.
Then $-Q^{\dagger}\Phi = \Psi$ , i.e. \quad\quad\quad $iq_a\phi_a \ = \ \psi$
\begin{equation}
  iq_c\phi_{abc} \ + \ 4 \ \overline{\partial}_{[a}\phi_{b]} \ = \ \psi_{ab}
\end{equation}
\centerline\vdots

$iq_{a_{2k+1}}\phi_{a{_1}}\cdots_{a{_{2k+1}}} \ + \ (2k)
2\overline{\partial}_{[_{a{_1}}} \phi_{a{_2}}\cdots_{a{_{2k}}]} \ = \
\psi_{a{_1}}\cdots_{a{_{2k}}}$ 

\centerline\vdots

$\psi_{a{_1}}\cdots_{a{_{2k}}} = (2K) 2 \overline{\partial}_[{a{_1}}
\phi_{a{_2}}\cdots_{a{_{2k}}]}$.

\vspace{0.5cm}
While for the verification of the first $(K-1)$ equations, in (14), it is
sufficient to only use (6) (and (12)) the last equality requires knowledge of
how $\psi_{a{_1}}\cdots_{a{_{2k}}}$ is solved in terms of
$\psi_{a{_1}}\cdots_{a{_{2k-2}}}$, i.e. (9)$_{k=K}$. While the square
integrability of $\Psi$ presumably implies $\| q\Phi \| < \infty$ (via (13)),
the discussion of whether $Q^{\dagger}\Psi=0$ actually implies  $\| \Phi \| <
\infty$ (when $\Phi$ is defined via (13)) will be more complicated
(but could perhaps roughly go as follows: $Q^{\dagger}\Psi=0$, resp. $Q\Psi=0$,
implies that the first derivatives of 
the $\Psi^{(n)}$, divided by $q$, are square-integrable, which is in
conflict with the normalizability of $\Psi$, unless $\Psi(q=0)=0$ -- which
probably improves the behavior of $\Phi$ at $q=0$ such that $\| \Phi \|
< \infty$, when $\| q\Phi \| < \infty$).

If $\Psi=Q^{\dagger}\Phi$, with $\| \Phi \| < \infty$, $\Psi$ can not be
annihilated by $Q$, due to the direct sum decomposition of ${\cal
  H}_+$ into $Q{\cal H}_{-}$, $Q^{\dagger}{\cal H}_{-}$, and states
annihilated by both $Q$ and $Q^{\dagger}$ (I am grateful to J.~Fr\"ohlich for
pointing out to me this simple but important fact, which suggested to check
whether the solutions (7)-(11) arise as images of $Q$, resp. $Q^{\dagger}$).

\vfill\eject
Consider now the general case,
\begin{eqnarray}
Q_{\beta} & = & D_a^{(\beta)}\partial_{\lambda{_a}} \ + \
M_a^{(\beta)}\lambda_a \nonumber\\
Q_{\beta}^{\dagger} & = & D_a^{(\beta)\dagger}\lambda_a \ + \
 M_a^{(\beta)\dagger}\partial_{\lambda_a}
\end{eqnarray}
In $D=11$ $(\Gamma^{j}=\Gamma^{j\dagger} =  -\Gamma^{jtr} , \ j = 1\cdots 7)$
or $D=4 (\Gamma^j  \to  0 , \ x_j \to 0):$
\begin{eqnarray}
D_{\alpha A}^{(\beta)} & = &
\delta_{\alpha\beta} \ 2\partial_A-if_{ABC} \ x_{jB}\overline{z}_C\Gamma^j_{\alpha\beta}\\
M_{\alpha A}^{(\beta)} & = & \delta_{\alpha\beta} \
iq_{A}+i\Gamma_{\alpha\beta}^j 
\frac{\partial}{\partial x_{jA}} - \frac{1}{2} 
f_{ABC}x_{jB}x_{kc}\Gamma_{\alpha\beta}^{jk} \ , \nonumber
\end{eqnarray}
where $A = 1\cdots N^{2}-1, \alpha\beta = 1\cdots 8 \ (D=11) \Gamma^{jk} = \frac{1}{2} \ [\Gamma_j,\Gamma_k]$. 
They satisfy [7]
\begin{equation}
\{Q_{\beta}, Q_{\beta'} \}  =  2i
  \delta_{\beta\beta'}\overline{z}_EJ_E 
\quad\quad \{Q_{\beta}, Q_{\beta'}^{\dagger}\} = 
\delta_{\beta\beta'}H \ + \ 2 \Gamma_{\beta\beta'}^j x_{jE}J_E
\end{equation}
with
\begin{equation}
J_E = -if_{EAA'}(x_{jA}\partial
x_{jA'}+z_A\partial_{A'}+\overline{z}_A\overline{\partial}_{A'}+\lambda_{\alpha
  A}\partial_{\lambda_{\alpha A'}}) \ = \ L_E + S_E
\end{equation}
and
\begin{equation}
H  =  (-\triangle + V) - 2if_{EAA'}x_{jE}\Gamma_{\alpha\alpha
  '}^j\lambda_{\alpha A}\partial_{\lambda_{\alpha ' A'}} +
f_{EAA'}z_E\lambda_{\alpha A}\lambda_{\alpha A'} +
f_{EAA'}z_E\partial_{\lambda_{\alpha A'}}\partial_{\lambda_{\alpha A}} 
\end{equation}
where $\triangle= 4\partial_A\overline{\partial}_A +
\partial_{jA}\partial_{jA}$ and $V=q^2 + \tilde{V}$ being twice
$V(x,\frac{1}{\sqrt{2}} Z, \frac{1}{\sqrt{2}}\overline{Z})$ given in
e.g. (4.11) of [7].
The superalgebra (15) alone (!) implies the following (commutation) relations
($(\beta\beta ')$ denoting symmetrisation $(\frac{1}{2}\beta\beta ' +
\frac{1}{2} \beta '\beta)$):
\begin{eqnarray}
\left[D_a^{(\beta}, D_{a'}^{\beta ')} \right]& = & 0 \\
\left[ M_a^{(\beta}, M_{a'}^{\beta ')} \right] & = & 0 \\
D_a^{(\beta}M_a^{\beta ')} & = & i \delta_{\beta\beta '}\overline{z}_EL_E \\
\left[ M_{\alpha A}^{(\beta}, D_{\alpha ' A'}^{\beta ')} \right] & = &
\delta_{\alpha\alpha 
  '}\delta_{\beta\beta '}\overline{z}_Ef_{EAA'}
\end{eqnarray}
and, using also the specific form of $H$, (19),
\begin{eqnarray}
\left[ M_{[\alpha A}^{\beta}, D_{\alpha ' A']}^{\dagger\beta '} \right] & = &
\delta_{\alpha\alpha 
  '}\delta_{\beta\beta '}z_Ef_{EAA'}\\
D_a^{\beta}D_a^{\beta '\dagger} + M_a^{\dagger\beta '}M_a^{\beta} & = &
\delta_{\beta\beta '}(-\triangle + V) + 2\Gamma_{\beta\beta '}^j x_{jE}L_E
\end{eqnarray}
\begin{equation}
\left[ M_{\alpha A}^{\beta}, M_{\alpha ' A'}^{\beta '\dagger} \right] + \left[
  D_{\alpha A}^{\beta ' \dagger}, D_{\alpha 'A'}^{\beta} \right] =
-2ix_{jE}f_{EAA'} (\delta_{\beta\beta '}\Gamma_{\alpha\alpha
  '}^j+\delta_{\alpha\alpha '}\Gamma_{\beta\beta '}^j) \ .
\end{equation}
The equations $Q^{(\beta)}\Psi=0, Q^{(\beta)\dagger}\Psi =0$,
read:
\begin{eqnarray}
(2k-1) M^{(\beta)}_{[a_1}\psi_{a{_2}}\cdots_{a{_{2k-1}}}] & = &
D_{a{_{2k}}}^{(\beta)}\psi_{a{_1}}\cdots_{a{_{2k}}}\\
k & = & 1, \cdots , K \nonumber\\
(2k-1) D^{(\beta)\dagger}_{[a_1}\psi_{a{_2}}\cdots_{a{_{2k-1}}}] & = &
M_{a{_{2k}}}^{(\beta)\dagger}\psi_{a{_1}}\cdots_{a{_{2k}}} \ .
\end{eqnarray}
Due to the non-commutativity of $M$ with $M^{\dagger}$ they are slightly more
difficult to solve (but also less singular, as $\vec{M}^{\dagger}\vec{M} >
0$).
The solution of the first equations,
$M_a\psi  =  D_b\psi_{ab}, D_a^{\dagger}\psi = M_b^{\dagger}\psi_{ab}$, are
$\psi = (M^{\dagger}M)^{-1}M_a^{\dagger}D_b\psi_{ab}$ and (with $\psi_{abc}$,
totally anitsymmetric, arbitrary)
\begin{equation}
\psi_{ab} = 2 D_{[a}^{\dagger}M_{b]}(M^{\dagger}M)^{-1}\psi +
M_c^{\dagger}\psi_{abc} \ .
\end{equation}

\vspace{2cm}
\centerline{\large Acknowledgement}
\vspace{0.5cm}
\noindent I would like to thank G.~Felder, J.~Fr\"ohlich, G.-M.~Graf,
H.~Nicolai and R.~Suter for valuable discussion, the Albert Einstein
Institute, Potsdam, and the Institute for Theoretical Physics, Z\"urich, for
hospitality, and the Deutsche Forschungsgemeinschaft for financial support.

\vspace{1cm}

\centerline{\large References}
\vspace{0.5cm}

\begin{description}
\item[[1]] M.~Porrati, A.~Rozenberg; hep-th/9708119.
\vspace{-0.3cm}
\item[[2]] S.~Sethi, M.~Stern; hep-th/9705046.
\vspace{-0.3cm}
\item[[3]] P.~Yi; hep-th/9704098.
\vspace{-0.3cm}
\item[[4]] J.~Fr\"ohlich, J.~Hoppe; hep-th/9701119.
\vspace{-0.3cm}
\item[[5]] T.~Banks, W.~Fischler, S.H.~Shenker, L.~Susskind; hep-th/9610043.
\vspace{-0.3cm}
\item[[6]] B.~de Wit, M.~L\"uscher, H.~Nicolai; Nuclear Physics B320 (1989)
  135.
\vspace{-0.3cm}
\item[[7]]  B.~de Wit, J.~Hoppe, H.~Nicolai; Nuclear Physics B305 (1988) 545.
\vspace{-0.3cm}
\item[[8]] R.~Flume; Annals of Physics 164 (1985) 189. \\
M.~Claudson, M.~Halpern; Nuclear Physics B 250 (1985) 689.\\
M.~Baake, P.~Reinicke, V.~Rittenberg; Journal of Math. Physics 26 1985) 1070.
\vspace{-0.3cm}
\item[[9]] J.~Hoppe; ``Quantum Theory of a Massless Relativistic Surface'',\\
  MIT Ph.D. Thesis 1982.\\ 
J.~Goldstone; unpublished.
\end{description}

\end{document}